\documentclass[preprint,5p,times,twocolumn]{elsarticle}
\usepackage{amssymb}
\usepackage{graphicx}
\usepackage{booktabs}
\usepackage{lineno}

\journal{Nuclear Instruments and Methods in Research A}

\newcommand{\degrees}{\ensuremath{^\circ} }

\makeatletter
\def\@author#1{\g@addto@macro\elsauthors{\normalsize%
    \def\baselinestretch{1}%
    \upshape\authorsep#1\unskip\textsuperscript{%
      \ifx\@fnmark\@empty\else\unskip\sep\@fnmark\let\sep=,\fi
      \ifx\@corref\@empty\else\unskip\sep\@corref\let\sep=,\fi
      }%
    \def\authorsep{\unskip,\space}%
    \global\let\@fnmark\@empty
    \global\let\@corref\@empty  %% Added
    \global\let\sep\@empty}%
    \@eadauthor={#1}
}
\makeatother

\begin{document}
\begin{frontmatter}

\author{J.D. Maxwell\corref{cor1}}
\ead{jdmax@mit.edu}
\author{C.S. Epstein}
\author{R.G. Milner}

\address{Laboratory for Nuclear Science, Massachusetts Institute of Technology, Cambridge, MA 02139 USA}
\cortext[cor1]{Corresponding author}

%\title{Liquid Crystal Retarder Based Discharge Polarimeter for Metastability Exchange Optical Pumping of $^3$He}
\title{Liquid Crystal Polarimetry for Metastability Exchange Optical Pumping of $^3$He}

\begin{abstract}
We detail the design and operation of a compact, discharge light polarimeter for metastability exchange optical pumping of $^3$He gas near 1\,torr under a low magnetic field. The nuclear polarization of $^3$He can be discerned from its electron polarization, measured via the circular polarization of 668\,nm discharge light from an RF excitation. This apparatus  measures the circular polarization of this very dim discharge light using a nematic liquid crystal wave retarder (LCR) and a high-gain, transimpedance amplified Si photodiode.  We outline corrections required in such a measurement, and discuss contributions to its systematic error.
\end{abstract}


\end{frontmatter}

%\linenumbers 

\section{Introduction}

Metastability exchange optical pumping (MEOP) allows the polarization of helium 3 nuclei in a low pressure gas using a uniform magnetic field and circularly polarized light~\cite{colgrove}. An electric discharge in the gas promotes a small fraction of the atoms into the 2S, metastable state. Transitions from the 2$^3$S$_1$ into the 2$^3$P$_0$ states are driven using 1083\,nm laser light, which will change the magnetic quantum number by $\pm 1$ depending on the circular polarization of the light. The polarization induced in the metastable population in this way is then transferred into the ground state population of the gas via metastability exchange collisions. 

By observing the circular polarization of the 668\,nm light given off by 3$^1$D$_2$ to 2$^1$P$_1$ transitions in the discharge~\cite{pavlovic}, we obtain a measure of the electron polarization, and thence the nuclear polarization after applying a correction determined by NMR calibrations. This correction is dependent on the pressure of the gas; as listed in reference~\cite{lorenzon}, the ratio of the nuclear polarization to electron polarization is 8.37 in $^3$He at 1.010\,torr.

\begin{figure}
\begin{center}
\includegraphics[width=3.2in]{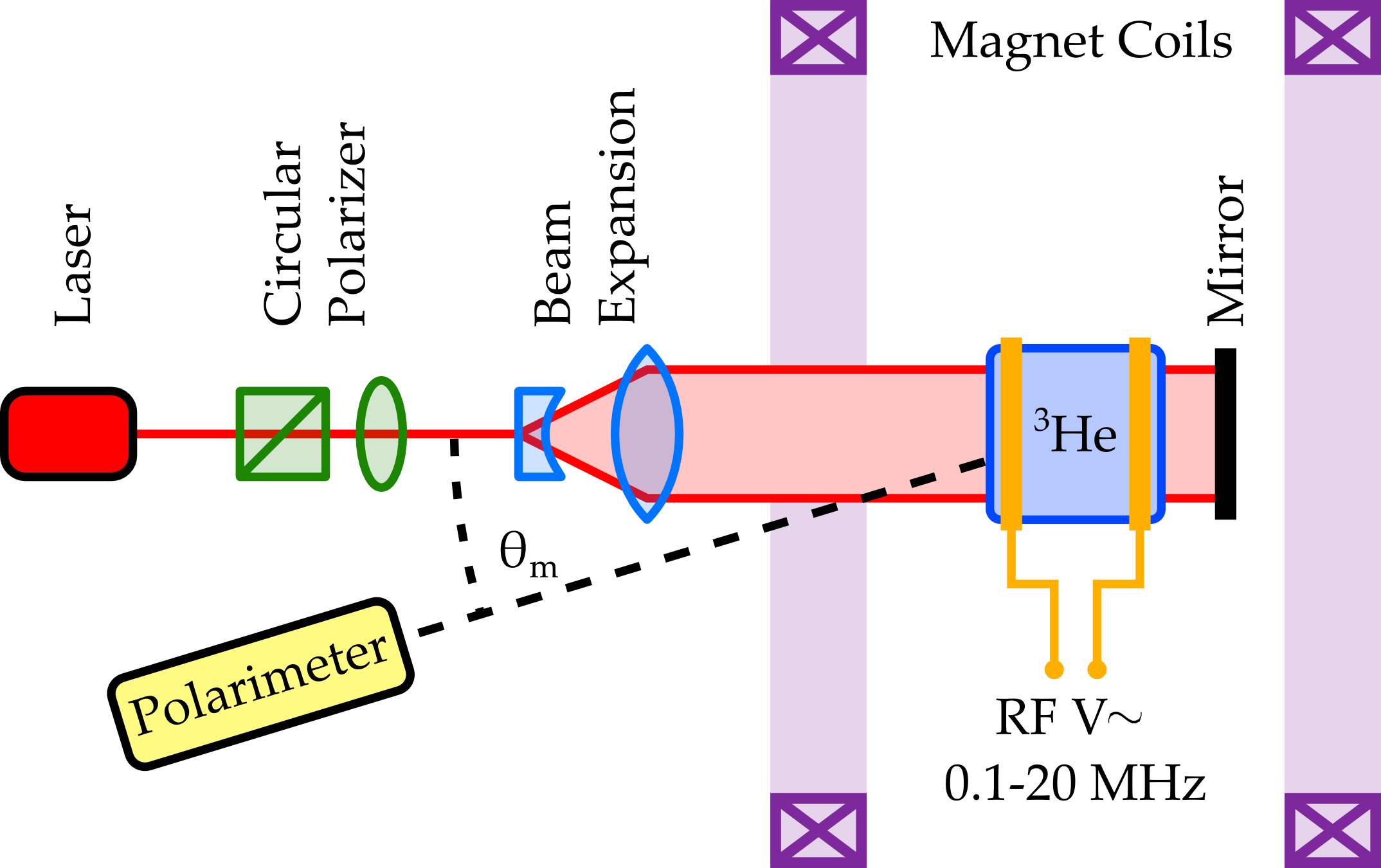}
\end{center}
\caption{Diagram of $^3$He polarizer and polarimeter setup.}
\label{setup}
\end{figure}

Figure \ref{setup} shows a simplified diagram of our polarizing apparatus. 1083\,nm laser light is circularly polarized and expanded to illuminate the volume of a glass pumping cell which is in a uniform 30\,G magnetic field. A dielectric mirror reflects 99.6\% of the laser light to increase pumping efficiency, while reflecting only 20\% of 668\,nm light, reducing the amount of reflected background light in the polarimeter. 
The polarimeter views the RF discharge in the gas cell at an angle $\theta_m$ to the holding field and at a distance of approximately 50\,cm. As the intensity of the RF discharge directly affects the rate of polarization and its maximum value, the polarimeter must be sensitive to very low light intensities viewed from this distance. 

\subsection{Measurement Techniques}

The degree of circular polarization of the discharge light, $P_d$, can be observed by sampling the intensity of left ($\sigma_-$) and right ($\sigma_+$) hand circularly polarized components of the light:

\begin{equation}
\label{eq:pol}
P_d = \frac{I(\sigma_+)-I(\sigma_-)}{I(\sigma_+)+I(\sigma_-)}.
\end{equation}

One approach to this measurement consists of a rotating quarter-wave plate and linear polarizer in series. As the slow axis of the quarter-wave plate reaches 45\degrees and 135\degrees to the axis of the linear polarizer, the left and right hand components are observed, creating an oscillating light intensity at twice the rotation frequency. The light polarization is then calculated as the peak to peak oscillation voltage as measured in a photodiode, divided by twice the RMS voltage.

This measurement can also be realized using two photodiodes, a linearly polarizing beam splitter cube and a quarter-wave plate at 45\degrees to the cube~\cite{stoltz}. The paths through the quarter-wave plate and subsequently transmitted or reflected in the beam-splitter are sensitive to left or right handed polarized light. The difference over the sum of the photocurrents in the detectors in these two paths give a measure of the polarization, though any difference in the efficiency or gain of the photodiodes introduces error in the measurement.

%In practice, this is done by processing the photomultiplier signal with a lock-in amplifier tuned to twice the frequency of rotation. The AC magnitude from the lock-in is then divided by the DC offset to produce the circular polarization. The setup is necessarily large, as the rotation mount and motor, the photomultiplier and its high-voltage power supply, each take up a good deal of space. As the space inside our polarizer's magnetic shielding is limited, it was important to reduce the size, without cost to the accuracy of the polarimeter. % Our rotating wave-plate polarimeter left room for improvement, as systematic uncertainty 

\section{Liquid Crystal Retarder Polarimeter Design}

The advent of variable waveplate retarders using nematic liquid crystals offers a new alternative in this context, although such retarders have been in use for at least 10 years in Mueller-matrix polarimeters~\cite{bueno,drouillard,daniels}. Right and left hand polarized components are accessed directly by changing the wave retardance from $\lambda/4$ to $3\lambda/4$ with the primary axis at 45\degrees to a linear polarizer.

\begin{figure}
\begin{center}
\includegraphics[width=3in]{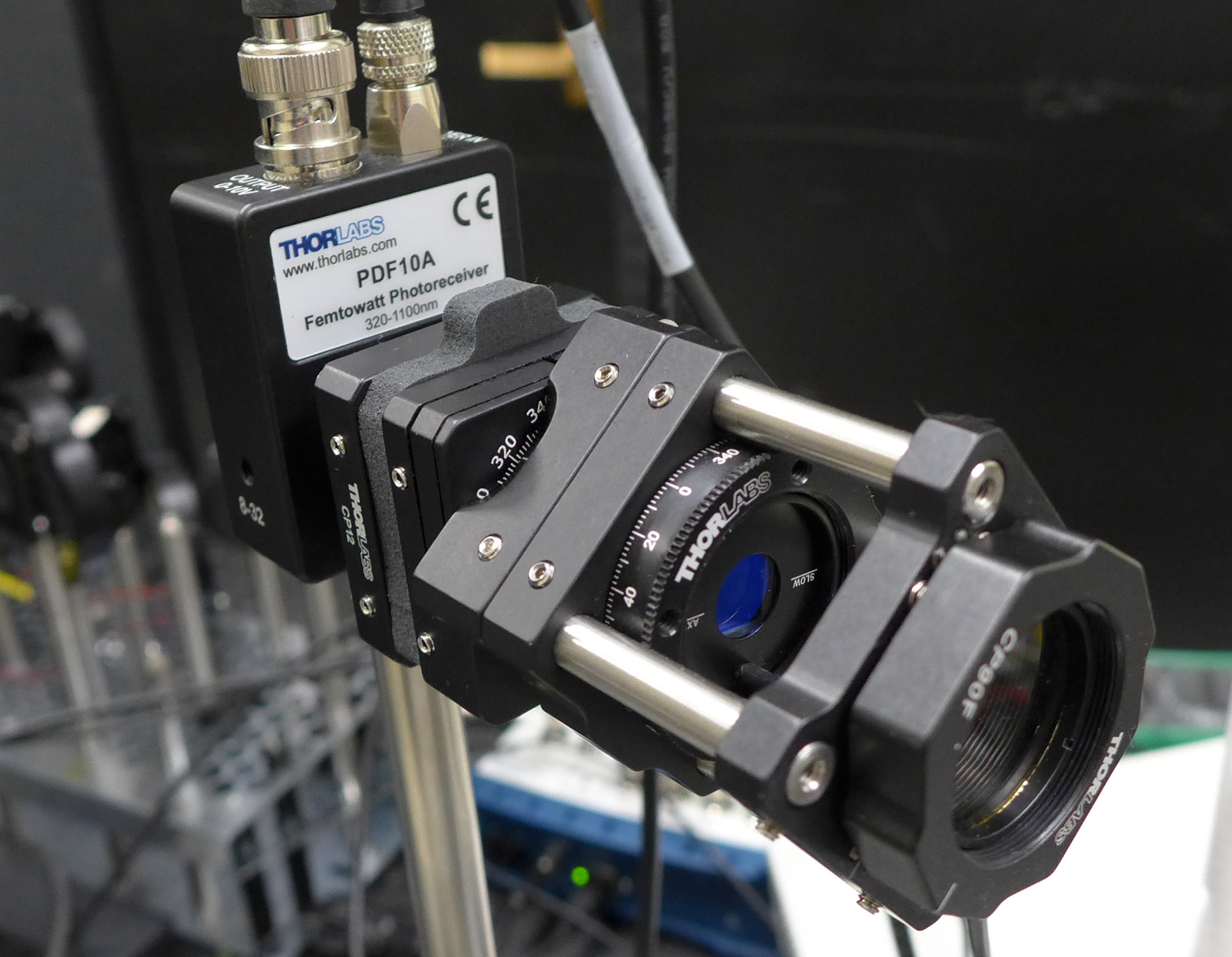}\\
\vspace{0.2in}
\includegraphics[width=3.3in]{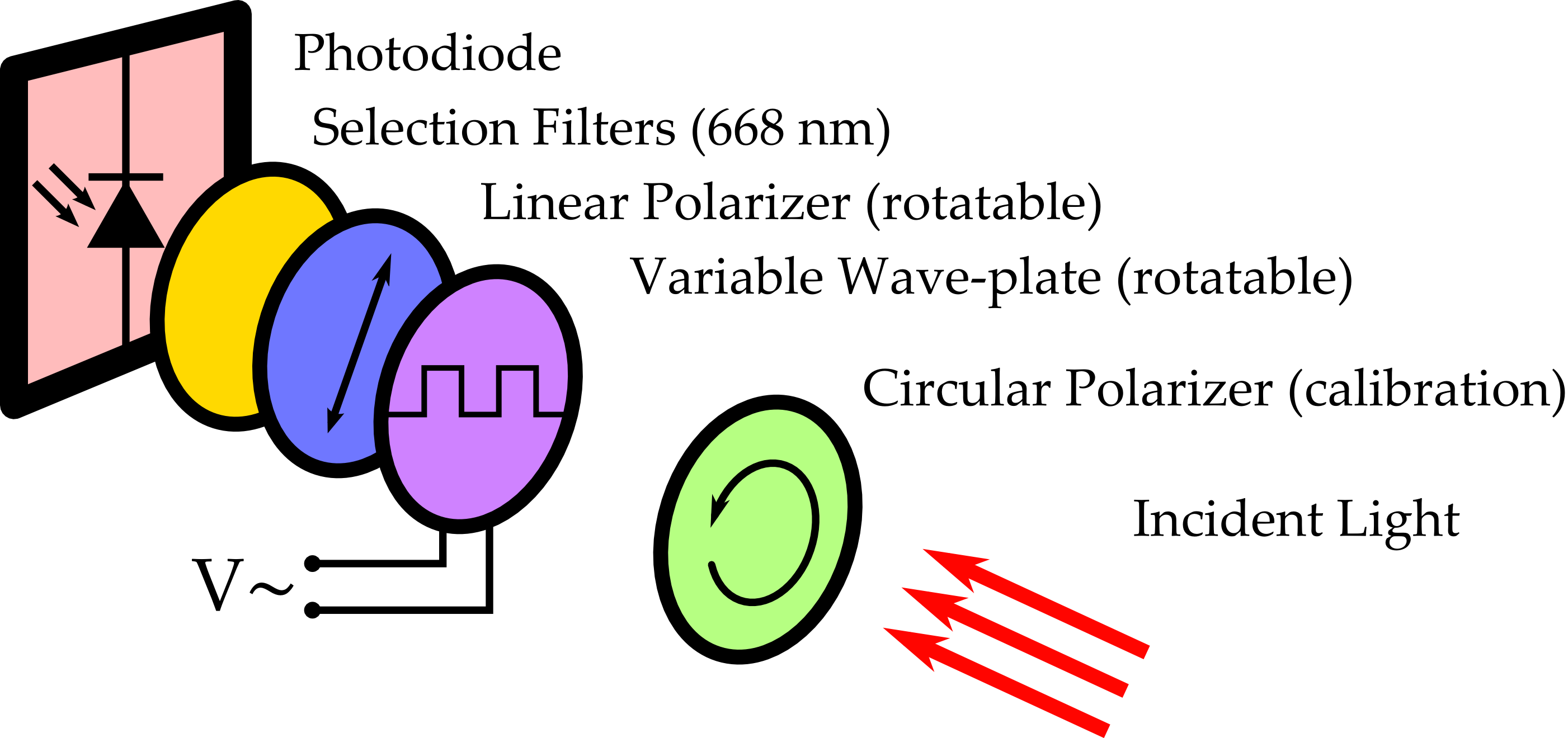}
\end{center}
\caption{Photograph of discharge polarimeter and diagram of its components.}
\label{fig:new_pol}
\end{figure}

The new polarimeter, shown in figure \ref{fig:new_pol}, was designed to be built from off-the-shelf parts and consists of 3 main components: a liquid crystal retarder (LCR), a linear polarizing filter and a photodiode. The linear polarizer is a Thorlabs dichroic film polarizer that provides approximately 9000:1 extinction ratio at 668\,nm. The retarder and polarizer are fitted in 2 rotation mounts so that they may rotate as an assembly and so that the variable waveplate may rotate in relation to this assembly for calibration.  The photodiode is fitted with an optics tube containing several filters to exclude background light. A removable circularly polarizing filter can be placed in front of the device to allow calibration.

% temperature dependence

\subsection{Liquid Crystal Retarder}

Our LCR is a Thorlabs LCC1112-A, which allows variable retardance from 0 to greater than 3/4 wave, and is anti-reflective coated from 350 to 700\,nm. %\cite{lcr}
 The oblong shape and ordered structure of nematic crystal molecules in the crystal create an optical anisotropy. By aligning the molecules using an electric voltage, the birefringence of the crystal can be tuned. 

In practice, the retardance is held constant by sweeping the voltage rapidly from positive to negative to avoid damaging the crystal with ionic buildup. The Thorlabs LCC25 liquid crystal controller was used to generate the required 2\,kHz supply voltage. In addition to aligning the liquid crystal with an oscillating supply voltage, this controller also allows switching between two RMS voltages to correspond to each desired wave retardance. These set voltages can be chosen either by internal clock or external TTL trigger.

A key limitation of liquid crystal retarders is the switching speed, which is asymmetric as the crystals switch between high and low absolute voltage. The rise time from high retardance and low voltage to low retardance and high voltage is 0.5\,msec, while the return takes on the order of 15\,msec. These switching times are temperature dependent. In practice our measurements have been taken at roughly twice a second, where our chosen sampling time is much longer than the LCR switching time.  %This compares unfavorably with the near 200 Hz sample rates of the rotation based polarimeter, but the added time resolution is not needed with our polarization rates and the observed increase in precision for the new method is great. %The precision of the LCR 

Temperature dependence in liquid crystal retarders is caused by the simple change in density of the liquid crystal material. In practice, this effect is negligible after a short warm-up period, and recalibration of the LCR is a simple remedy to the effect. Temperature stabilized crystals are available, but not necessary in this application.

\subsection{Photodiode}

The discharge intensity in the $^3$He cell directly affects both the maximum achievable polarization and the rate of polarization. At a given pressure and excitation frequency, higher intensity generally results in higher pumping rate, but lower maximum polarization~\cite{gentile}. Ideally, a polarimeter should operate well in a wide range of brightness to accommodate either case. While decreasing the light intensity to avoid saturating the detector is accomplished simply with neutral density filters, a photodiode with too little sensitivity is more problematic. We thus erred on the side of over-sensitivity in choosing a candidate photodiode.

To this end, the Thorlabs PDF10A femtowatt receiver was chosen. The femtowatt receiver combines a low noise silicon photodiode with a very high gain amplifier, sacrificing output bandwidth for amplification. At 668\,nm, the silicon photodiode response is approximately 0.45\,A/W, while the gain factor is $1\times 10^9$\,kV/A. The maximum noise equivalent power is quoted by Thorlabs as $1.4 \times 10^{-15}$, which allows detection down to 10\,fW of optical power. The output voltage noise is typically 6.5\,mV RMS. While the bandwidth is limited below 20\,Hz, the time scales of interest are still easily resolved for this application. %\cite{pd} 
A neutral density filter or focusing lens can be included as needed to reduce or augment the photodiode voltage to ensure a favorable signal to noise without saturation of the data acquisition.

\subsection{Light Filtration}

As our photodiode is sensitive to a wide range of wavelengths, several filters are necessary select the 668\,nm light which corresponds to the transition of interest. Light from other sources in the lab, as well as discharge light not at the desired wavelength, creates a significant background, even with such filters in place. Reducing the 1083\,nm light from the pumping laser, which can be up to 10\,W, is of particular importance.

Previous experimenters who have used LCRs in this context have leveraged modulated RF discharge with a lock-in amplifier to isolate the portion of the photocurrent due to the gas excitation~\cite{stoltz}.  While this technique may be used to improve our methods, we find that light filtration and background subtraction were adequate to achieve the desired accuracy. In addition, the modulation of the discharge intensity requires a range of discharge intensity that would affect the degree and rate of polarization in the gas. % constrained by the noise floor of our photodiode package in our laboratory conditions.

Two bandpass and one shortpass filters are used for light filtration. The Thorlabs FB650-40 650 $\pm$ 40 \,nm bandpass filter transmits 0.72 of 668\,nm light and $7.7 \times 10^{-8}$ of 1083\,nm, and a second bandpass filter, the FB670-10 670 $\pm$ 2\,nm, further tightens the band of allowed light (0.57 at 668\,nm and $3 \times 10^{-5}$ at 1083\,nm). To still further reduce the background from the 10\,W, 1083\,nm pumping laser, Thorlabs FES0800 shortpass filter cuts the laser light to $5 \times 10^{-6}$, while transmitting 0.90 of 668\,nm light. %\cite{shortpass}
All three filters are housed in an optics tube mounted directly to the photodiode package. The combination of these filters doesn't provide a factor of $10^{-21}$ reduction at 1083\,nm in practice, as a significant background of laser light is seen as the laser power increases, as is discussed in section \ref{offset}. 

\section{Method and Corrections}

The $^3$He polarization as measured from the circular polarization of the discharge light is

\begin{equation}
\label{eq:meas}
P = \frac{f_{p}}{\cos\theta_m}\left( \frac{I_{3/4}-I_{1/4}}{I_{3/4}+I_{1/4}-2I_{b}}\right)
\end{equation}
for light intensities at the two waveplate settings $I_{3/4}$ and $I_{1/4}$, angle of polarimeter $\theta_m$ as in figure \ref{setup}, pressure calibration factor $f_{p}$, and background light intensity $I_{b}$. Prerequisite to this measurement are the calibration and alignment of the polarimeter, and the measurement of the background light intensity $I_{b}$.

\subsection{Calibration and Alignment}
\label{sec:cal}

To calibrate the device, the angle between the transmission axis of the linear polarizer and the slow axis of the LCR is set, and the voltages corresponding to quarter and three-quarter wave are set. To facilitate this calibration, an Edmunds plastic circular polarizing filter is placed in front of the polarimeter, as seen in figure \ref{fig:new_pol}. Once a rough alignment of the LCR to 45\degrees to the linear polarizer is performed, the voltage calibration of the LCR is performed.

\begin{figure}
\begin{center}
\includegraphics[width=3.2in]{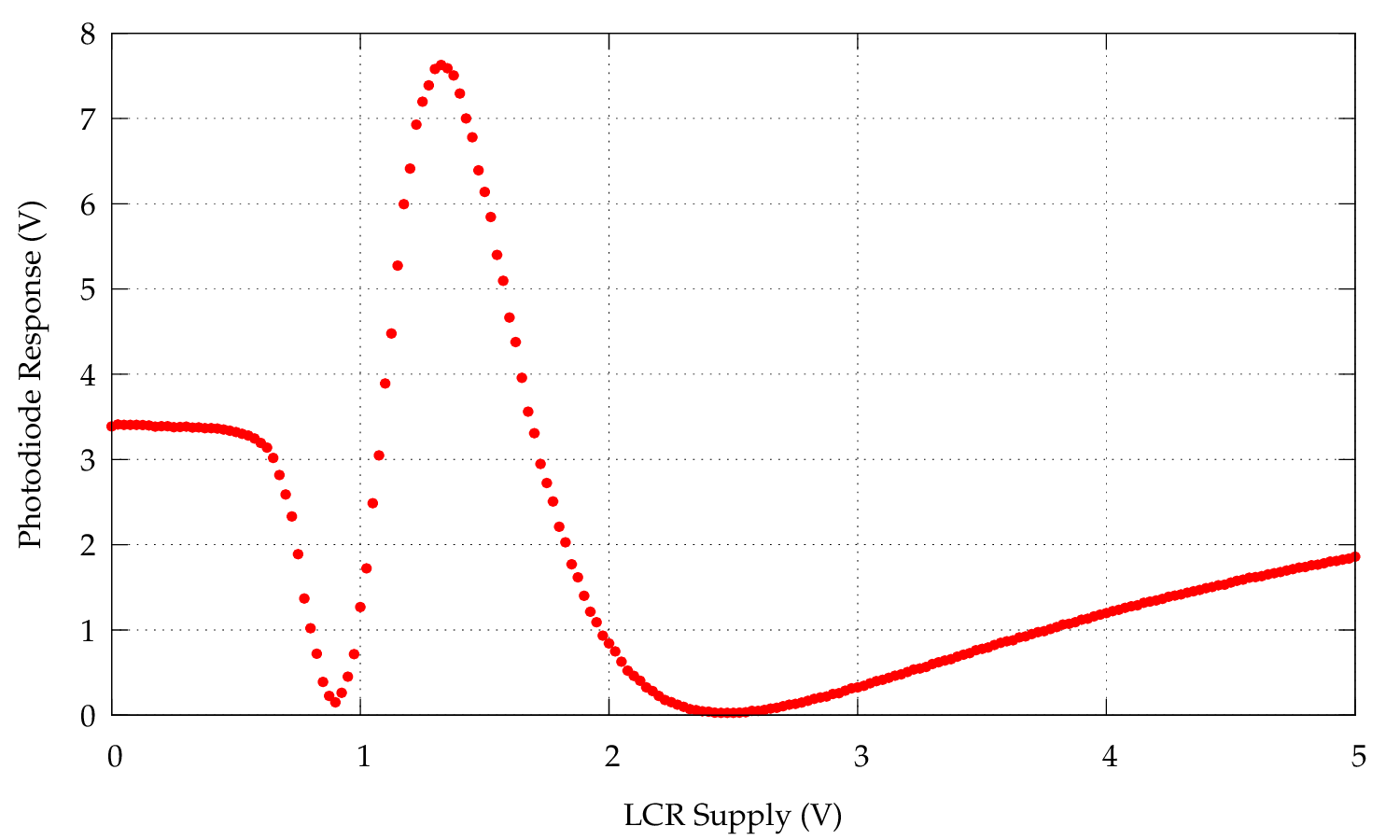}
\end{center}
\caption{Calibration curve from light intensity response versus LCR supply voltage with circular polarizer in place. Here $\lambda/4$ occurs near 2.5\,V, and $3\lambda/4$ near 1.3\,V. For this technique $5\lambda/4$ is not used, but occurs near 0.9\,V.}
\label{fig:response}
\end{figure}

By increasing the LCR supply voltage and monitoring the photodiode voltage, a response curve is produced, such as the one seen in figure \ref{fig:response}. At low voltage, the retardance is above $5\lambda/4$, and as the voltage is increased, the brightness reaches a minimum at $5\lambda/4$, a maximum at $3\lambda/4$, and a second minimum at $\lambda/4$. Voltages which produced the maximum and minimum response are set as the 1/4 and 3/4 wave points. 

With the desired LCR voltage points found, the polarimeter is set to switch between these points as the data acquisition system records the resulting photodiode voltage, producing a value for the circular polarization of the light. With the circular polarizer in place, any value away from 100\% indicates misalignment or other inaccuracy in the measurement. The alignment of the LCR to the linear polarizer is then adjusted to maximize the measured polarization at as close to 100\% as achievable.

Finally, the circular polarizer is removed and the LCR and linear polarizer assembly is rotated to remove the effects of linearly polarized light. Light from the discharge which reflects off a surface such as the mirror, or is transmitted through the glass, can become linearly polarized. Depending on the orientation of the LCR and linear polarizer in the polarimeter, this linear polarization can result in a false polarization signal. Fortunately both the S and P polarization from reflection and transmission occurs in a single plane in our setup, allowing this effect to be removed by rotating the LCR and linear polarizer until the circular polarization of the discharge light is zero with the polarizing magnetic field off.

\subsection{Intensity Measurement}

The measurement of the light intensity is performed through a National Instruments USB-6259 BNC data acquisition system. The signals with the LCR at $\lambda/4$ and $3\lambda/4$ retardance are observed to determine the switching response time of the waveplate, and a delay is chosen to ensure the waveplate is stable for each measurement. For each setting of the variable waveplate, the photodiode voltage is measured between 200 and 800 times at between 1 and 4\,kHz. The resulting points are averaged into one voltage for that waveplate setting, and the standard deviation of these points is produced to quantify high frequency noise or any other shifts in the time scale of the measurement. Samples with very large standard deviations can be discarded as erroneous, likely the result of changing ambient light. 

This method relies on LabView software control of the variable waveplate to directly obtain $I_{3/4}$ and $I_{1/4}$. In reference~\cite{kraft}, the waveplate is continuously switched at a given frequency, rather than switched in software. The polarization is then determined as the ratio of the  RMS photodiode voltage from a lock-in amplifier at the switching frequency over the mean photodiode voltage from an integrator. Although our method is necessarily slower, it avoids the need for extra amplification and integration instruments, and allows us to remove the switching response of the LCR directly.

\subsection{Background Subtraction}
\label{offset}
Background light that makes it through the filters results in a simple offset to the photodiode voltage which can be subtracted from the measured intensity $I_m$ as $I = I_m - I_b$. Unpolarized background light will result in the same intensity $I_b = I_b(3/4) = I_b(1/4)$ for both waveplate settings, resulting in the correction in equation \ref{eq:meas}. The photodiode voltage is measured with the discharge off to obtain $I_b$ both when the laser is on and off. Using static subtractions such as these makes the measurement vulnerable to changes in background light, such as power fluctuations from the pumping laser, so the background offsets are remeasured periodically.

Polarized background light will result in a falsely augmented circular polarization measurement. The final alignment step in section \ref{sec:cal} ensures that polarimeter is not sensitive to vertical linear polarization, either from transmission through the glass cell or from the pumping laser. The background light at both waveplate settings is compared to ensure they are equal and this false signal removed.

\subsection{Error}

From equation \ref{eq:meas}, the error in the nuclear polarization $\sigma_P$ is related to the measured light intensities $I_{3/4}$ and $I_{1/4}$ and their errors $\sigma_{3/4}$ and $\sigma_{3/4}$ as
\begin{equation}
\label{eq:err}
\sigma_P = \frac{2f_p}{\cos\theta_m(I_{3/4}+I_{1/4})^2}\sqrt{I_{3/4}^2\sigma_{1/4}^2+I_{1/4}^2\sigma_{3/4}^2}.
\end{equation}
 This relation makes clear the importance of maximizing the light intensities measured in the photodiode to improve the signal to noise, as the noise floor of the measurement does not improve. The error is inversely proportional to the signal voltage, so that the error in a 4\,V signal is twice as large of that at 8\,V.
By taking many points at each setting, we remove some electrical noise and leave the error of each point dependent on accuracy of our measurement devices. Typical noise for the femtowatt receiver is quoted at 6\,mV RMS, and the data acquisition board claims absolute accuracy of 1.9\,mV at full scale. We conservatively assume a combined systematic uncertainty from these devices of 0.01\,V, which is much larger than the standard error on the mean voltage for each setting in usable measurements.

Three further sources of systematic error can be minimized with care. Imperfection in the waveplates, as well as any mis-alignment, act to falsely decrease the measured polarization. This effect is apparent during calibration of the polarimeter, as shown in figure \ref{fig:response}; a non-zero minimum at $\lambda/4$ indicates imperfection in the waveplates, alignment or the circular polarizing film used during the calibration. Falsely increased polarization can come from linearly polarized light of certain orientation, and as mentioned in section \ref{sec:cal}, a measurement of zero polarization indicates this effect has been removed during the calibration. Finally, changes in ambient light or laser light intensity add further error, which is minimized by periodically resetting the offsets used to correct both effects. Changing ambient light levels on faster time scales result in rejected measurements due to high standard deviations. Based on measurements during calibration, we typically achieve less than 1.5\% additional absolute error from these sources. 

Table \ref{tab:err} summarizes typical contributions to the systematic error in the measured optical polarization. The combined error of 0.19\% absolute becomes 1.6\% absolute error in the helium polarization after applying the pressure calibration factor. The electron polarization to helium polarization pressure calibration from Lorenzon~\cite{lorenzon} contributes an additional systematic error in $P_n/P$ of 2.1\% relative at 1 torr.

\begin{table}
\begin{center}
\begin{tabular}{l c}
\toprule
Source of error  & Absolute \% \\
\midrule
Data Acquisition Board & 0.03\% \\
Photodiode & 0.06\% \\
Linearly polarized light & 0.1\% \\
Misalignment  & 0.1\% \\
Background subtraction & 0.1\% \\
\cmidrule(r){2-2}
Combined & 0.19\% \\
\bottomrule
\end{tabular}
\end{center}
\caption{Estimated contributions to systematic error on the optical polarization of the discharge light for a typical measurement, listed in percent absolute. After applying the pressure calibration, the combined error becomes approximately 1.6\% absolute in the helium polarization.}
\label{tab:err}
\end{table}

\begin{figure}
\begin{center}
\includegraphics[width=3.2in]{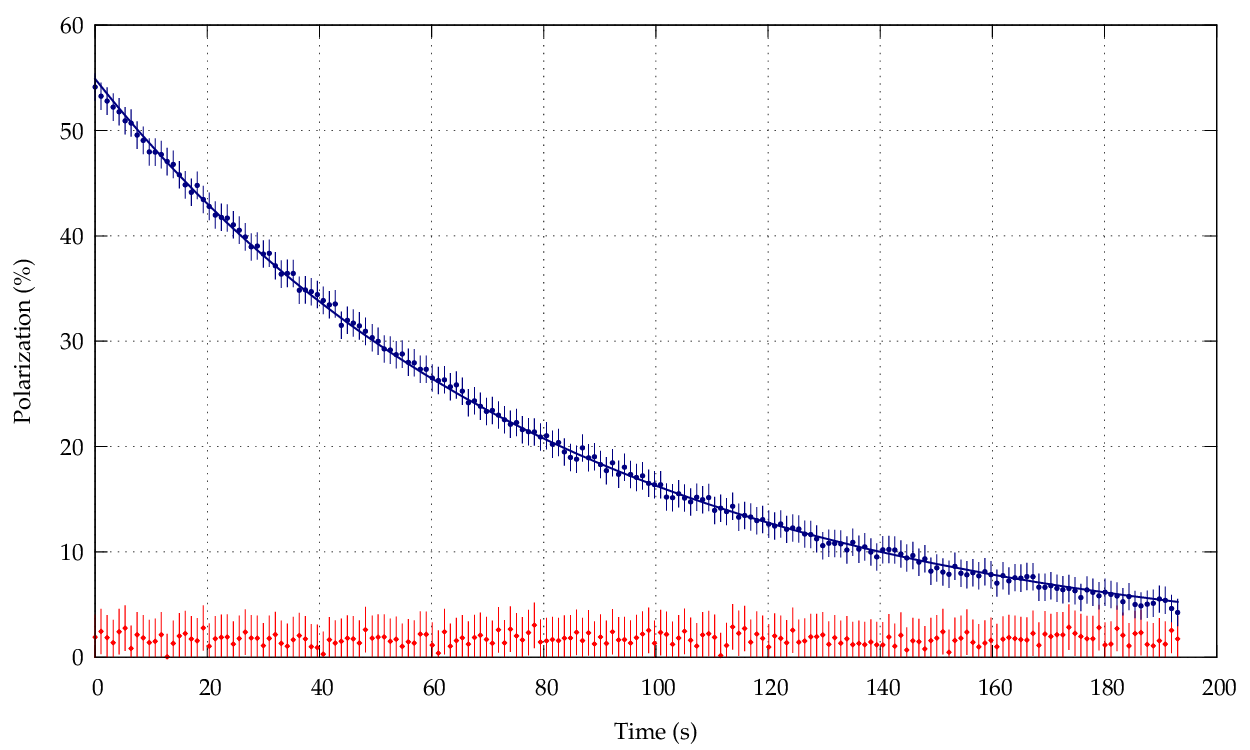}
\end{center}
\caption{The polarization over time in two helium cells as measured with two polarimeters, including statistical and estimated systematic errorbars, but excluding error from pressure calibration. Blue circles show a cell with decaying polarization, and red diamonds show a cell which remains unpolarized. }
\label{fig:err}
\end{figure}

Figure \ref{fig:err} shows the polarization measured simultaneously by two different polarimeters in two different 1\,torr $^3$He cells at 30\,G, where the errorbars include statistical and estimated systematic error. The plot illustrates both the quality of the method for polarization relaxation and the importance of maintaining high signal to noise. The polarization in one cell, denoted with blue circles, is decaying with a relaxation time of 82 seconds, and while the other, in red diamonds, is left unpolarized.  The decaying cell is measured with an advantageous signal to noise, with an 8.5\,V photodiode voltage at $3\lambda/4$ with 0.01\,V error results in 0.76\% absolute error for a typical point. In figure \ref{fig:err} these points are plotted with this measured error, as well as an additional 1\% systematic error from other sources. The cell at rest had a dimmer discharge and more background noise, so that the 4.1\,V photodiode voltage at $3\lambda/4$ results in 1.49\% absolute error for a typical point. In addition, linear polarization from refraction was not entirely removed in this case, resulting in a small offset from zero; in the figure, these points are plotted with an additional 1.5\% systematic error accordingly.

%We do not include this uncertainty in figure \ref{fig:err}. %This uncertainty thus dominates the systematic error for this measurement, discouraging further effort to tighten down the error from other sources.

\section{Conclusion}
We have produced reliable, accurate measurements of helium-3 polarization during MEOP without the need for discharge modulation using a device easily assembled from off-the-shelf parts. A high-gain photodiode allows the use of several optical filters to exclude background light while maintaining high signal to noise.  Background light which is not excluded by the filters is typically static and can be subtracted. Our measurements were performed at roughly once per second, which while more than adequate for our needs, could be improved significantly by reducing the sampling time at very little cost to accuracy. With proper calibration and background subtraction, a 1\% absolute measurement is feasible with this device.

\section*{Acknowledgments}
We gratefully acknowledge Andreas Kraft (now of Univ.~of Washington) for introducing us to the use of liquid crystal retarders in MEOP polarimeters at Mainz. This research was funded by the Office of Nuclear Physics in the U.S. Department of Energy under the program for Research and Development for Next Generation Nuclear Physics Accelerator Facilities. 

\bibliography{discharge_pol}
\bibliographystyle{elsarticle-num}

%\begin{thebibliography}{00}
%\bibitem{colgrove} Colegrove et al, \textit{Polarization of He$^3$ Gas by Optical Pumping,} Phys.\ Rev.\ 132 (1963) doi: \url{http://dx.doi.org/10.1103/PhysRev.132.2561}.
%\bibitem{pavlovic} Pavlovic, Laloe, \textit{Etude d'une nouvelle méthode permettant d'orienter, par pompage optique, des niveaux atomiques excités.} J. Phys, (Paris), 31, 173 (1970) doi: \url{http://dx.doi.org/10.1051/jphys:01970003102-3017300}.
%\bibitem{lorenzon} Lorenzon, Gentile, Gao, McKeown, \textit{NMR Calibration of optical measurement of nuclear polarization in $^3$He}. Phys.\ Rev.\ A 47 (1993) doi: \url{http://dx.doi.org/10.1103/PhysRevA.47.468}.
%\bibitem{bueno} Bueno, \textit{Polarimetry using liquid-crystal variable retarders: theory and calibration}, J. Opt. A: Pure Appl. Opt. 2, 2000 doi: \url{http://dx.doi.org/10.1088/1464-4258/2/3/308}.
%\bibitem{drouillard} T. Drouillard et al, \textit{Polarimetry Using Liquid Crystal Variable Retarders}, \url{http://www.meadowlark.com/store/applicationNotes/Polarimetry\%20using\%20liquid\%20crystal\%20variable\%20retarders.pdf}
%
%\bibitem{daniels} J.M. Daniels \textit{Determination of the Circular Polarized Fraction of Light using a Voltage Driven Nematic Liquid Crystal Wave-Plate}, Mol. Cryst. Liq. Cryst., Vol. 434, 2005 doi: \url{http://dx.doi.org/10.1080/15421400590954849}.
%
%\bibitem{lcr} Thorlabs Inc, \textit{Three-Quarter-Wave Liquid Crystal Variable Retarders},
% \url{http://www.thorlabs.com/NewGroupPage9.cfm?ObjectGroup_ID=6338}
%\bibitem{bandpass} Thorlabs Inc, \textit{Bandpass \& Laser Line Filters},
%\url{http://www.thorlabs.com/thorProduct.cfm?partNumber=FB670-10}
%
%\bibitem{shortpass} Thorlabs Inc, \textit{Edgepass Filters},
%\url{http://www.thorlabs.com/newgrouppage9.cfm?objectgroup_id=918}
%\bibitem{pd} Thorlabs Inc, \textit{Si Transimpedance Amplified Photodetectors},
%\url{http://www.thorlabs.com/newgrouppage9.cfm?objectgroup_id=3257}
%\end{thebibliography}
\end{document}